\documentstyle[aps,preprint,epsf]{revtex}

\tighten

\begin{document}
\draft

\title
{\bf Optical Conductivity in a Two -- Dimensional Model of the Pseudogap State}
\author{M.V.Sadovskii,\ N.A.Strigina}
\address
{Institute for Electrophysics,\ Russian Academy of Sciences,\ Ural Branch,\\ 
Ekaterinburg,\ 620016, Russia\\
E-mail:\ sadovski@iep.uran.ru,\ strigina@iep.uran.ru} 
\maketitle


\begin{abstract}
We consider a two -- dimensional model of the pseudogap state, based on the
scenario of strong electron scattering by fluctuations of ``dielectric''
(AFM, CDW) short -- range order. We construct a system of recurrence
equations both for one -- particle Green's function and vertex part,
describing electron interaction with an external field, which take into
account all Feynman graphs for electron scattering by short -- range order
fluctuations. The results of detailed calculations of optical conductivity
are presented for different geometries (topologies) of the Fermi surface,
demonstrating both the effects of pseudogap formation and 
localization effects. These results are in qualitative agreement with
experimental data obtained for high -- temperature superconducting cuprates.
\end{abstract} 
\pacs{PACS numbers:  74.25-q, 74.25.Jb, 74.72-h, 74.20.Mn}

\newpage
\narrowtext
\section{Introduction}
One of the main problems in the physics of high -- temperature
copper oxide superconductors (HTSC) remains the nature of the so called
pseudogap state \cite{Tim,MS}, existing in the wide region of the
phase diagram. Pseudogap anomalies in electronic properties of these
systems are observed in a number of experiments, such as measurements of
optical conductivity, NMR, inelastic neutron scattering and angle -- resolved
photoemission (ARPES) \cite{Tim,MS}. In our opinion \cite{MS}, 
the preferable scenario for the pseudogap formation in copper oxides is
based on the picture of strong scattering of current carriers on well
developed fluctuations of ``dielectric'' (antiferromagnetic (AFM) or
charge density wave (CDW)) short -- range order, existing in the 
appropriate region of the phase diagram. This scattering is strong
in the vicinity of a characteristic scattering vector
${\bf Q}=(\frac{\pi}{a},\frac{\pi}{a})$ ($a$ -- is the lattice spacing of the
two -- dimensional lattice), corresponding to the doubling of lattice
period (vector of antiferromagnetism) and is a ``precursor'' of the spectrum
transformation due to the appearance of the long -- range AFM -- order.
Accordingly, in this region of the phase diagram develops essentially
non -- Fermi liquid like renormalization of electronic spectrum in certain
parts of the momentum space, close to the ``hot spots'' on the Fermi
surface \cite{MS}, where this surface is effectively ``destroyed''.
Direct experimental confirmation of this scenario for the pseudogap
formation was obtained in recent ARPES -- experiments on
$Nd_{1.85}Ce_{0.15}CuO_4$ \cite{Arm}, where the pseudogap renormalization of
electronic spectrum was observed in the vicinity of {\em separate} 
``hot spots''.

Within this scenario of the pseudogap formation we can formulate a
simplified ``nearly exactly'' solvable model, describing the main
anomalies of this state \cite{MS}, which takes into account all Feynman
diagrams of perturbation theory for scattering by (Gaussian) fluctuations
of short -- range order with characteristic scattering vector from the
vicinity of ${\bf Q}$, determined by the appropriate correlation length
$\xi$ \cite{Sch,KS}. This model is based on the two -- dimensional
generalization of the pseudogap formation in one -- dimension due to
fluctuations of CDW -- type short -- range order proposed
rather long ago by one of the authors \cite{S74,S79}. Simplified variant of
this two -- dimensional model (``hot patches'' model) was used in
Refs. \cite{SP,KS1,KS2,KS3} to describe basic anomalies of superconducting
state, formed on the ``background'' of the ``dielectric'' pseudogap.

In Refs. \cite{Sch,KS} mainly one -- particle properties of the model were
analyzed, such as the spectral density and the density of states.
The remarkable property of this model is the possibility to sum also the whole
Feynman series for the two -- particle problem for the vertex part,
describing the system response to an external field (e.g. electromagnetic)
\cite{S74,S91,ST91}. In the simplified version  of the ``hot patches'' model 
such calculations for the two -- dimensional case were performed in
Ref. \cite{S99}. The main aim of the present work is the detailed analysis of
both the theoretical aspects of calculations of two -- particle properties
of the more general model \cite{Sch,KS} and calculations of the optical
conductivity for different geometries (topologies) of the Fermi surface,
appearing for ``realistic'' enough spectrum of non -- interacting
electrons.

\section{``Hot spots'' model.}

\subsubsection{Description of the model and ``nearly exact'' solution for the
one -- particle Green's function.}

\begin{figure}
\epsfxsize=16cm
\epsfysize=20cm
\epsfbox{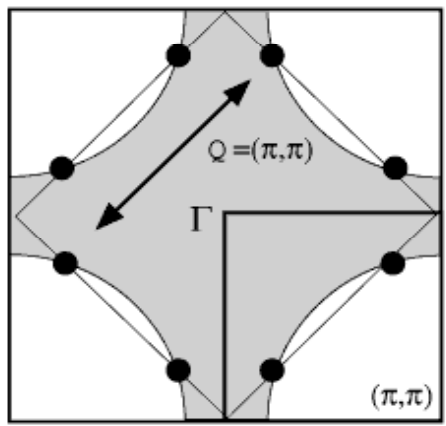}
\caption{Fermi surface with ``hot spots'', connected by the scattering 
vector of the order of ${\bf Q}=(\frac{\pi}{a},\frac{\pi}{a})$.}
\label{hspts}
\end{figure}

In the model of ``nearly antiferromagnetic'' Fermi -- liquid, which is
actively used to explain the microscopic nature of HTSC \cite{MBP,MB},
effective electron interaction with spin -- fluctuations of AFM short --
range order is usually described by the dynamic spin susceptibility
$\chi_{\bf q}(\omega)$ of the following form, determined by fitting to NMR
data \cite{MB}:
\begin{equation}
V_{eff}({\bf q},\omega)=g^2\chi_{\bf q}(\omega)\approx
\frac{g^2\xi^2}{1+\xi^2({\bf q-Q})^2-i\frac{\omega}{\omega_{sf}}}
\label{V}
\end{equation}
where $g$ -- is a coupling constant, $\xi$ -- correlation length of spin
fluctuations, ${\bf Q}=(\pi/a,\pi/a)$ -- vector of antiferromagnetic ordering
in dielectric phase, $\omega_{sf}$ -- characteristic frequency of spin
fluctuations. Both dynamic susceptibility and effective interaction
(\ref{V}) have a sharp maximum in the region of  ${\bf q}\sim {\bf Q})$, 
leading to two ``types'' of quasiparticles --- ``hot'' one with momenta in
the vicinity of ``hot spots'' on the Fermi surface
(Fig. \ref{hspts}) and ``cold'' with momenta around the parts of the Fermi
surface around Brillouin zone diagonals \cite{Sch}. This is due to the fact
that quasiparticles from the regions around the
``hot spots'' are strongly scattered with the momentum transfer of the order
of ${\bf Q}$ by spin fluctuations (\ref{V}), while for quasiparticles with
momenta far from the ``hot spots'' this interaction is small enough.

For high enough temperatures  $\pi T\gg \omega_{sf}$ we can neglect the
spin dynamics  \cite{Sch}, and use the static approximation in (\ref{V}):
\begin{equation}
V_{eff}({\bf q})=\tilde\Delta^2\frac{\xi^2}{1+\xi^2({\bf q-Q})^2}
\label{Vef}
\end{equation}
where $\tilde\Delta$ -- is an effective parameter with dimension of energy.
In the following, similar to Refs. \cite{Sch,KS}, we consider both
$\tilde\Delta$ and $\xi$ as phenomenological parameters, to be determined from
experiments. Considerable simplification of calculations, allowing to analyze
higher order contributions of perturbation theory, can be achieved if we
replace (\ref{Vef}) by the model interaction of the following form \cite{KS}:
\begin{equation}
V_{eff}({\bf q})=\Delta^2\frac{2\xi^{-1}}{\xi^{-2}+(q_x-Q_x)^2}
\frac{2\xi^{-1}}{\xi^{-2}+(q_y-Q_y)^2}
\label{Veff}
\end{equation}
where $\Delta^2=\tilde\Delta^2/4$. Eq. (\ref{Veff}) is qualitatively the
same as (\ref{Vef}) and quantitatively is very close to it in most
interesting region of $|{\bf q-Q}|<\xi^{-1}$, determining the scattering in
the vicinity of ``hot spots''.

We assume the following spectrum of free (``bare'') quasiparticles \cite{Sch}:
\begin{equation}
\xi_{\bf p}=-2t(\cos p_xa+\cos p_ya)-4t^{'}\cos p_xa\cos p_ya
 - \mu
\label{spectr}
\end{equation}
where $t$ -- is the transfer integral between nearest neighbors, while
$t'$ -- is the transfer integral for second nearest neighbors on the
square lattice, $\mu$ -- is the chemical potential. This expression gives
rather good approximation to the results of band structure calculations for
real HTSC -- systems, e.g. for  $YBa_2Cu_3O_{6+\delta}$ we have $t=0.25eV$,\ 
$t'=-0.45t$ \cite{Sch}. Chemical potential $\mu$ is fixed by carrier 
concentration. In this paper we shall consider different characteristic
values of $t$, $t'$ and $\mu$, leading to different geometries (topologies) of
Fermi surfaces, with no relation to any particular system, with the aim of
analyzing general enough picture.

Consider first -- order in $V_{eff}$ correction to electronic self -- energy,
determined by the simplest Feynman diagram ($\varepsilon_n=(2n+1)\pi T$):
\begin{equation}
\Sigma(\varepsilon_n{\bf p})=\sum_{\bf q}V_{eff}({\bf q})
\frac{1}{i\varepsilon_{n}-\xi_{\bf p+q}}
\label{sig}
\end{equation}
For large enough correlation lengths $\xi$, the main contribution to the
sum over ${\bf q}$ comes from the region close to ${\bf Q}=(\pi/a,\pi/a)$. 
Then we can write:
\begin{equation}
\xi_{\bf p+q}=\xi_{\bf p+Q+k}\approx \xi_{\bf p+Q}+{\bf v_{\bf p+Q}k}
\label{linsp}
\end{equation}
where ${\bf v}_{\bf p+Q}=\frac{\partial\xi_{\bf p+Q}}{\partial
{\bf p}}$ -- is the appropriate velocity of the quasiparticle on the
Fermi surface. Then (\ref{sig}) is easily calculated to be:
\begin{equation}
\Sigma(\varepsilon_n{\bf p})=\frac{\Delta^2}{i\varepsilon_n-\xi_{\bf p+Q}
+i(|v^x_{\bf p+Q}|+|v^y_{\bf p+Q}|)\kappa sign\varepsilon_n}
\label{sigm}
\end{equation}
where $\kappa=\xi^{-1}$. Let us stress that ``linearization'' of quasiparticle
spectrum (\ref{linsp}) in (\ref{sig}) is done only for a small 
(of the order of $v_F\xi^{-1}$), due to a large values of $\xi$, correction
to an energy spectrum of quasiparticles close to the Fermi surface, while
the form of the spectra $\xi_{\bf p}$ and $\xi_{\bf p+Q}$ is given
by general expression (\ref{spectr}).

In Ref. \cite{KS} we have performed a detailed analysis of higher -- order
contributions for $\Sigma(\varepsilon_n{\bf p})$. It was shown that in case of
equal signs of velocity projections (on the Fermi surface)
$v^x_{\bf p}$ and $v^x_{\bf p+Q}$, as well as of $v^y_{\bf p}$ and 
$v^y_{\bf p+Q}$, Feynman integrals for any diagram of any order are
determined only by contributions from the poles of Lorentzians in
(\ref{Veff}) and can be easily calculated 
\footnote{Analogous statement is valid also for a special case of
velocities in ``hot spots'', connected by vector ${\bf Q}$, being exactly
perpendicular to each other \cite{Sch}.}. In this case the contribution of an
arbitrary diagram of the $N$-th order over effective interaction
(\ref{Veff}) is given by:
\begin{equation}
\Sigma^{(N)}(\varepsilon_n{\bf p})=\Delta^{2N}\prod_{j=1}^{2N-1}
\frac{1}{i\varepsilon_n-\xi_{j}({\bf p})+in_jv_j\kappa}
\label{Ansatz}
\end{equation}
where $\xi_j({\bf p})=\xi_{\bf p+Q}$ and 
$v_j=|v_{\bf p+Q}^{x}|+|v_{\bf p+Q}^{y}|$ 
for odd $j$, while $\xi_j({\bf p})=\xi_{\bf p}$ and $v_{j}=
|v_{\bf p}^x|+|v_{\bf p}^{y}|$ for even $j$. 
Here $n_j$ -- is the number interaction lines surrounding $j$-th
Green's function in a given diagram and for definiteness we assumed 
$\varepsilon_n>0$.
  
In Ref. \cite{KS} we have given a detailed analysis of conditions for
these requirements for velocities in the points on the Fermi surface,
connected by vector ${\bf Q}$ (``hot spots'') to be valid, and also presented
some examples of appropriate geometries of the Fermi surface, which are
realized for certain relations between transfer integrals
$t$ and $t'$ in Eq. (\ref{spectr}). Under these conditions expression
(\ref{Ansatz}) is exact, with only limitation due to the use of abovementioned
``linearization'' in Eq. (\ref{linsp}). In general case (for other
relations between $t$ and $t'$) we use (\ref{Ansatz}) as a fortunate
{\em Ansatz} for the contribution of an arbitrary order, obtained by simple
continuation over spectrum parameters $t$ and $t'$ to the region of interest.
Even in the worst one -- dimensional case \cite{S79}, equivalent to the
case of the simple square Fermi surface, appearing in two -- dimensions
for (\ref{spectr}) with $t'=0$ and $\mu=0$, the use of this {\em  Ansatz}
gives results (e.g. for the density of states) which are quantitatively very
close \cite{Sad} to the results of an exact numerical simulation of this
problem \cite{Brt}. In this sense we are using the term ``nearly exact''
solution.

In case of the use of this {\em Ansatz} Eq. (\ref{Ansatz}) it is easily seen,
that a contribution of some arbitrary diagram with crossing interaction
lines is in fact equal to the contribution of some definite diagram of the
same order without intersections of interaction lines \cite{S79}. Thus we can
limit ourselves only to diagrams without crossing interaction lines, taking
those with intersections into account with the help of additional
combinatorial factors, attributed to ``initial'' interaction vertices
(or interaction lines) \cite{S79}. As a result we obtain the following
recursion relation for the one -- electron Green's function (continuous
fraction representation) \cite{S79}), giving an effective algorithm for
numerical computations \cite{KS}:
\begin{eqnarray}
G_{k}(\varepsilon_n\xi_{\bf p})=\frac{1}{i\varepsilon_n-
\xi_k({\bf p})+ikv_k\kappa
-\Sigma_{k+1}(\varepsilon_n\xi_{\bf p})}\equiv\nonumber\\
\equiv\left\{G^{-1}_{0k}(\varepsilon_n\xi_{\bf p})
-\Sigma_{k+1}(\varepsilon_n\xi_{\bf p})\right\}^{-1}
\label{G}
\end{eqnarray}
\begin{equation}
\Sigma_{k}(\varepsilon_n\xi_{\bf p})=\Delta^2\frac{v(k)}
{i\varepsilon_n-\xi_k({\bf p})+ikv_k\kappa-
\Sigma_{k+1}(\varepsilon_n\xi_{\bf p})}
\label{rec}
\end{equation}
In graphic form this recursion relation for the Green's function is shown
in Fig. \ref{recurr}. The physical Green's function of interest to us is
given by $G(\varepsilon_n\xi_{\bf p})=G_{k=0}(\varepsilon_n\xi_{\bf p})$.  
In (\ref{G}) we also introduced a helpful technical notation: 
\begin{equation} G_{0k}(\varepsilon_n\xi_{\bf p})=
\frac{1}{i\varepsilon_n-\xi_k({\bf p})+ ikv_k\kappa} 
\label{G0k} 
\end{equation}
Combinatorial factor:
\begin{equation}
v(k)=k
\label{vcomm}
\end{equation}
for our case of commensurate fluctuations with
${\bf Q}=(\pi/a,\pi/a)$ \cite{S79}, if we neglect the spin variables
(CDW -- fluctuations). If we take into account the spin structure of
interaction within the model of ``nearly antiferromagnetic'' Fermi liquid
(spin -- fermion model of Ref. \cite{Sch}), combinatorics of diagrams
becomes more complicated. In this case, spin -- conserving scattering
formally gives commensurate combinatorics, while spin -- flip scattering
is described by diagrams with incommensurate combinatorics
(``charged'' random field in terms of Ref. \cite{Sch}). As a result, the
recursion relation for the Green's function (\ref{rec}) is still valid, 
but with different combinatorial factor $v(k)$ \cite{Sch}:  
\begin{equation} 
v(k)=\left\{\begin{array}{cc}
\frac{k+2}{3} & \mbox{for odd $k$} \\
\frac{k}{3} & \mbox{for even $k$}
\end{array} \right.
\label{vspin}
\end{equation}
Below we limit ourselves to the cases of (\ref{vcomm}) and (\ref{vspin}),
some details concerning the case of incommensurate CDW -- fluctuations
can be found in Ref. \cite{KS,S74,S79}.

Our solution for one -- particle Green's function is exact in the limit of
$\xi\to\infty$, when it can be found also in analytic form \cite{S74,Sch}.
It is also exact in a trivial limit of $\xi\to 0$, when for fixed value of
$\Delta$ interaction (\ref{Veff}) just vanishes. For all intermediate values
of $\xi$ it gives apparently very good interpolation, being practically
exact for certain geometries of the Fermi surface, which are realized for
definite regions of spectrum parameters in (\ref{spectr}) \cite{KS}.

Using (\ref{G}) we can easily perform numerical computations of
one -- electron spectral density:  
\begin{equation}
A(E{\bf p})=-\frac{1}{\pi}ImG^R(E{\bf p})
\label{spdns}
\end{equation}
which can be determined also from ARPES -- experiments \cite{MS}.
In (\ref{spdns}) $G^R(E{\bf p})$ denotes the retarded Green's function,
which is obtained by the usual analytic continuation of (\ref{G}) from
Matsubara's frequencies to the real axis of $E$. Similarly, we can perform
calculations of one -- electron density of states:
\begin{equation}
N(E)=\sum_{\bf p}A(E{\bf p})=-\frac{1}{\pi}\sum_{\bf p}ImG^R(E{\bf p})
\label{NE}
\end{equation}
Details of these calculations and discussion of results for our
two -- dimensional model can be found in Refs. \cite{Sch,KS}.

\subsubsection{Recurrence relations for the vertex part and conductivity.}

\begin{figure}
\epsfxsize=14cm
\epsfysize=20cm
\epsfbox{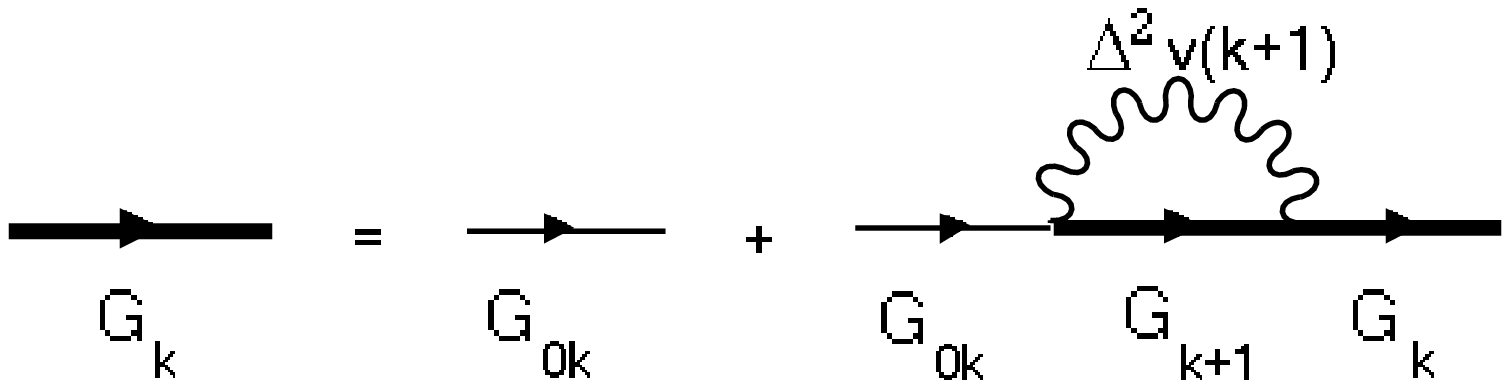}
\caption{Diagrammatic form of the recurrence relation for the Green's
function.}
\label{recurr}
\end{figure}

\begin{figure}
\epsfxsize=14cm
\epsfysize=20cm
\epsfbox{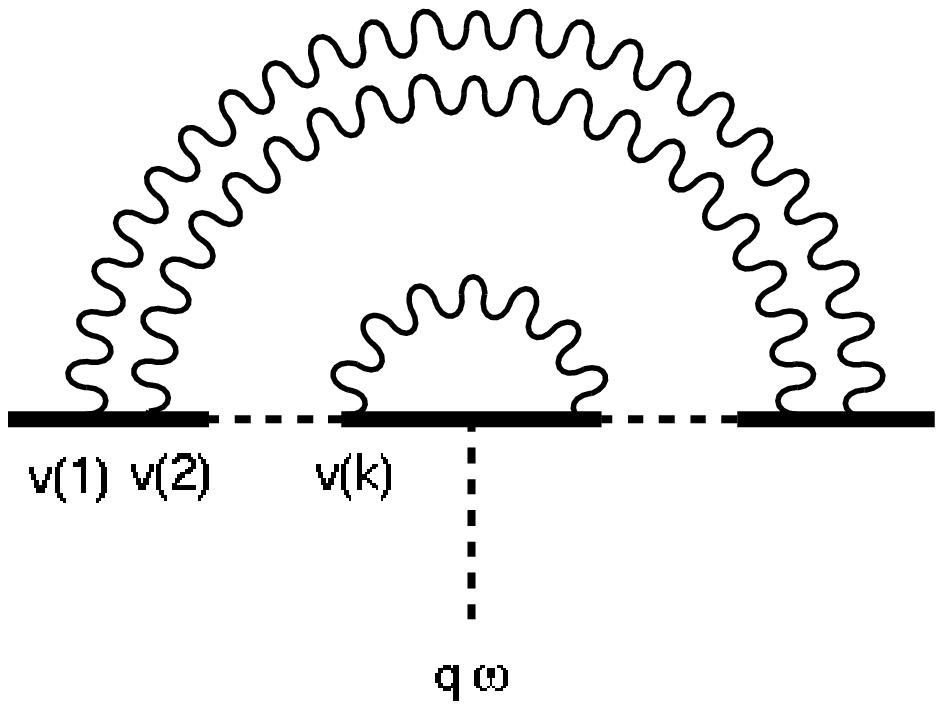}
\caption{General diagram for the vertex correction.}
\label{vert}
\end{figure}

\begin{figure}
\epsfxsize=14cm
\epsfysize=20cm
\epsfbox{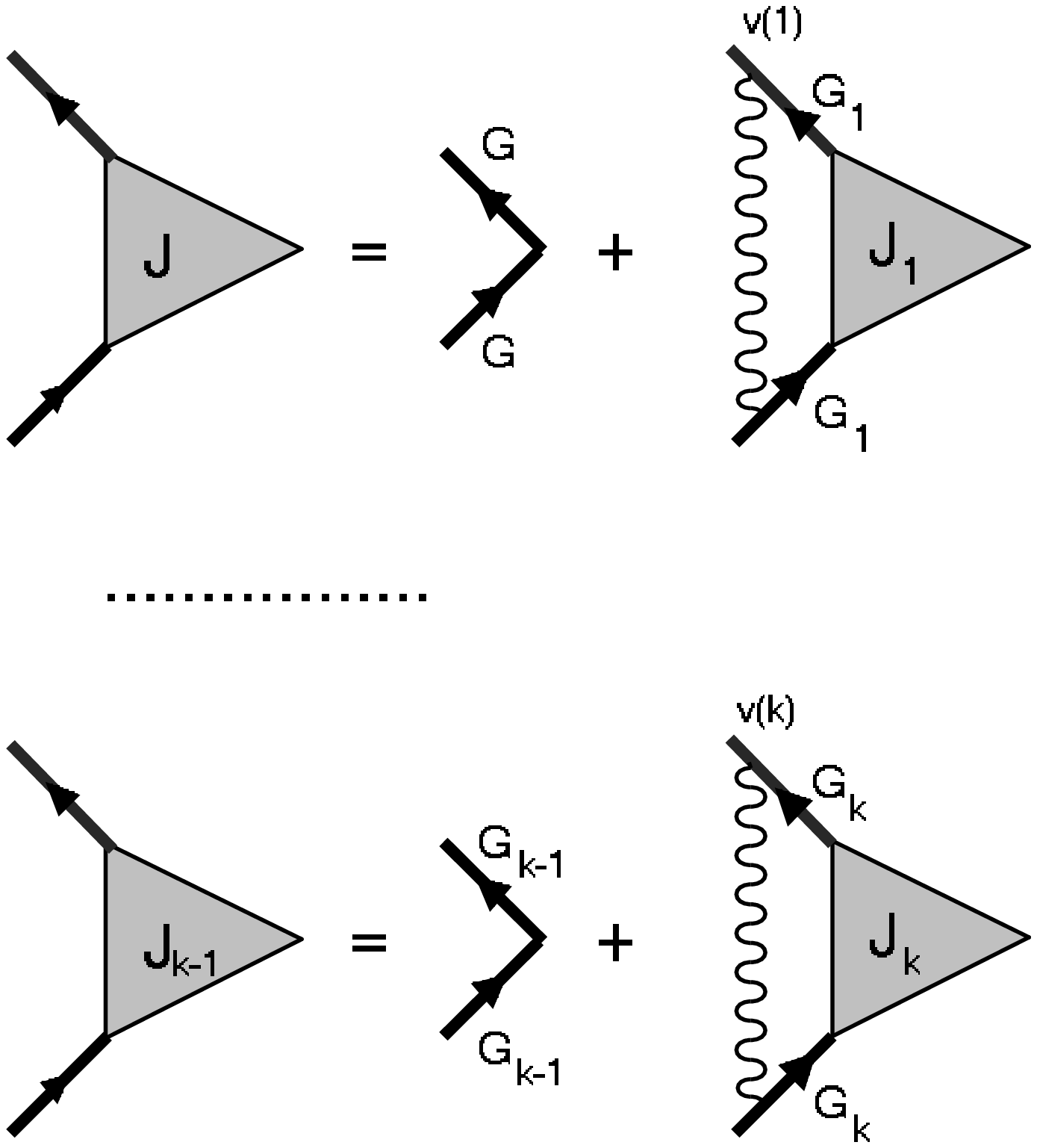}
\caption{Recurrence relations for the vertex part.}
\label{recvertx}
\end{figure}

\begin{figure}
\epsfxsize=14cm
\epsfysize=20cm
\epsfbox{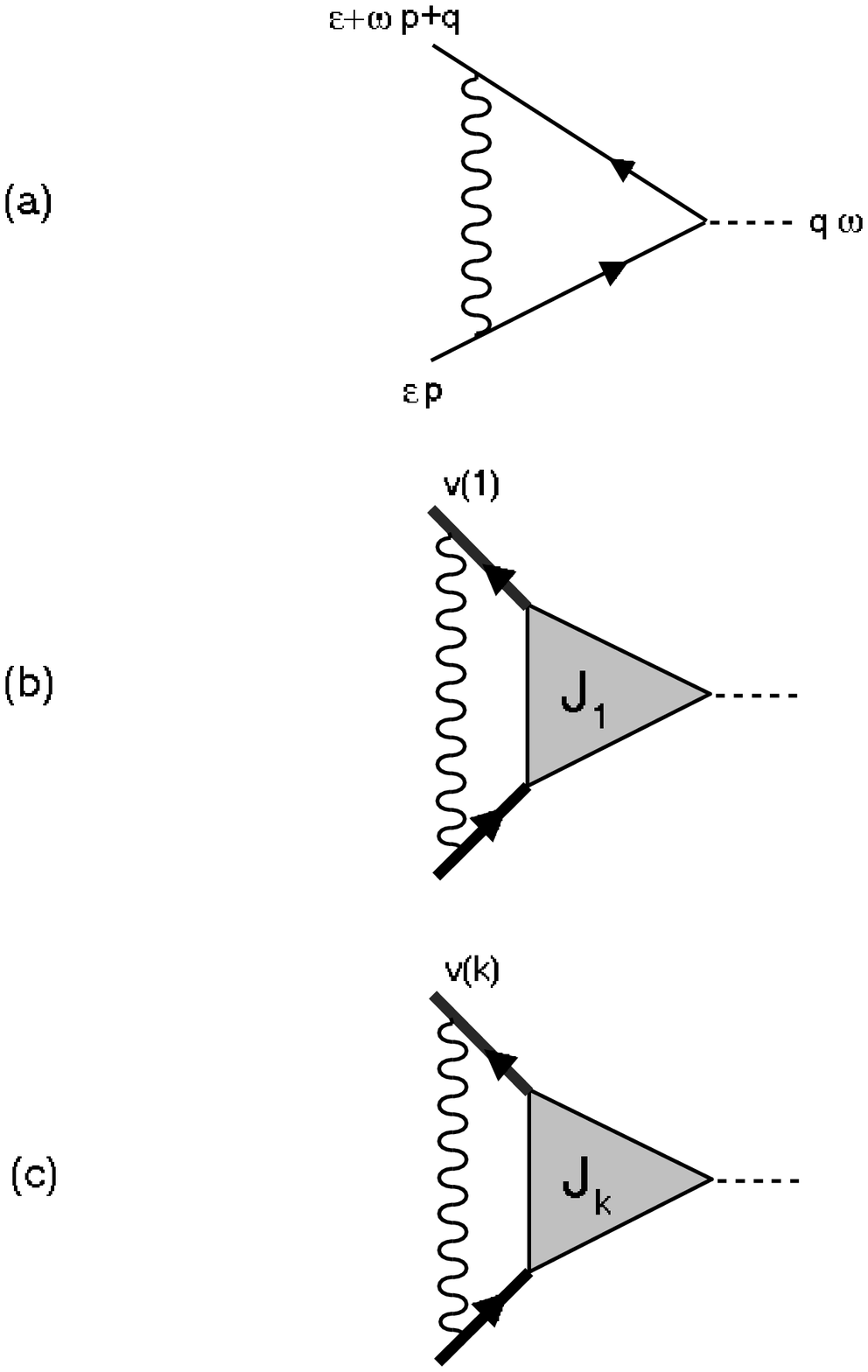}
\caption{Simplest corrections for vertex parts.}
\label{vertcorr}
\end{figure}

To calculate optical conductivity we need the knowledge of the vertex part
describing electromagnetic response of the system.
This vertex can be determined using the method suggested for the similar
one -- dimensional problem in Refs. \cite{S91,ST91}. Arbitrary diagram for the
vertex part can be obtained by an insertion of an external field line to the
appropriate diagram for the self -- energy \cite{S74}. We have already noted
that in our model we can limit ourselves only to diagrams with non --
intersecting interaction lines with additional combinatorial factors $v(k)$ 
in ``initial'' interaction vertices. It is clear then that to calculate vertex 
corrections we have to consider only diagrams of the type shown in 
Fig. \ref{vert}. Then we immediately obtain the system of recurrence equations
for the vertex parts shown by diagrams of Fig. \ref{recvertx}. To find 
appropriate analytic expressions consider the simplest vertex correction shown
in Fig. \ref{vertcorr} (a). Performing explicit calculations for $T=0$ 
in $RA$ -- channel we find its contribution to be:
\begin{eqnarray}
{\cal J}_1^{(1)RA}(\varepsilon{\bf p};\varepsilon+\omega{\bf p+q})=
\sum_{\bf K}V_{eff}({\bf K})G_{00}^A(\varepsilon\xi_{\bf p-K})
G_{00}^R(\varepsilon+\omega\xi_{\bf p-K+q})=\nonumber\\
=\Delta^2\left\{G_{00}^A(\varepsilon,\xi_1({\bf p})+iv_1\kappa)-
G_{00}^R(\varepsilon+\omega,\xi_1({\bf p+q})-iv_1\kappa)\right\}  
\frac{1}{\omega+\xi_1({\bf p})-\xi_1({\bf p+q})}=\nonumber\\
=\Delta^2G_{00}^A(\varepsilon,\xi_1({\bf p})+iv_1\kappa)
G_{00}^R(\varepsilon+\omega,\xi_1({\bf p+q})-iv_1\kappa)
\left\{1+\frac{2iv_1\kappa}{\omega+\xi_1({\bf p})-\xi_1({\bf p+q})}\right\}
\equiv
\nonumber\\
\equiv\Delta^2G_{01}^A(\varepsilon,\xi_{\bf p})
G_{01}^R(\varepsilon+\omega,\xi_{\bf p+q})
\left\{1+\frac{2iv_1\kappa}{\omega+\xi_1({\bf p})-\xi_1({\bf p+q})}\right\}
\label{J10}
\end{eqnarray}
where during the integral calculations we have used the following identity,
valid for the free -- electron Green's functions:
\begin{equation}
G_{00}^A(\varepsilon\xi_{\bf p})G_{00}^R(\varepsilon+\omega\xi_{\bf p+q})=
\left\{G_{00}^A(\varepsilon\xi_{\bf p})-G_{00}^R(\varepsilon+
\omega\xi_{\bf p+q})\right\}\frac{1}{\omega-\xi_{\bf p+q}+\xi_{\bf p}}
\label{ident0}
\end{equation}
``Dressing'' the internal electronic lines we obtain the diagram shown in Fig. 
\ref{vertcorr} (b), so that using the identity:
\begin{eqnarray}
G^A(\varepsilon\xi_{\bf p})G^R(\varepsilon+\omega\xi_{\bf p+q})=
\left\{G^A(\varepsilon\xi_{\bf p})-G^R(\varepsilon+\omega
\xi_{\bf p+q})\right\}\times\nonumber\\
\times\frac{1}{\omega-\xi_{\bf p+q}+\xi_{\bf p}
-\Sigma^R_1(\varepsilon+\omega\xi_{\bf p+q})+\Sigma^A_1(\varepsilon\xi_{\bf p})}
\label{ident} 
\end{eqnarray} 
valid for exact Green's functions, we can write the contribution of this
diagram as:
\begin{eqnarray}
{\cal J}_1^{RA}(\varepsilon{\bf p};\varepsilon+\omega{\bf p+q})=
\Delta^2v(1)G_{1}^A(\varepsilon,\xi_{\bf p})
G_{1}^R(\varepsilon+\omega,\xi_{\bf p+q})\times\nonumber\\
\times\left\{1+\frac{2iv_1\kappa}{\omega-\xi_1({\bf p+q})+\xi_1({\bf p})
-\Sigma_2^R(\varepsilon+\omega\xi_{\bf p+q})
+\Sigma^A_2(\varepsilon\xi_{\bf p})} 
\right\}J_1^{RA}(\varepsilon{\bf p};\varepsilon+\omega{\bf p+q}) 
\label{J1} 
\end{eqnarray}
Here we have assumed that interaction line in the vertex correction of 
Fig. \ref{vertcorr} (b) ``transforms'' self -- energies $\Sigma_1^{R,A}$ 
of internal lines into $\Sigma_2^{R,A}$,in accordance with our main
approximation for the self -- energy (cf. Fig. \ref{recurr})
\footnote{One of the main motivations for this trick is that it guarantees the 
fulfillment of an exact Ward identity to be discussed below.}

Now we can write down the similar expression for the general diagram,
shown in Fig. \ref{vertcorr} (c):
\begin{eqnarray}
{\cal J}_k^{RA}(\varepsilon{\bf p};\varepsilon+\omega{\bf p+q})=
\Delta^2v(k)G_{k}^A(\varepsilon,\xi_{\bf p})
G_{k}^R(\varepsilon+\omega,\xi_{\bf p+q})\times\nonumber\\
\times\left\{1+\frac{2iv_k\kappa k}{\omega-\xi_k({\bf p+q})+\xi_k({\bf p})
-\Sigma_{k+1}^R(\varepsilon+\omega\xi_{\bf p+q})
+\Sigma^A_{k+1}(\varepsilon\xi_{\bf p})} 
\right\}J_k^{RA}(\varepsilon{\bf p};\varepsilon+\omega{\bf p+q}) 
\label{Jk} 
\end{eqnarray}
Then we can write the general recurrence relation for the vertex part 
Fig. \ref{recurr} in the following form:
\begin{eqnarray}
J_{k-1}^{RA}(\varepsilon{\bf p};\varepsilon+\omega{\bf p+q})=
1+\Delta^2v(k)G_{k}^A(\varepsilon,\xi_{\bf p})
G_{k}^R(\varepsilon+\omega,\xi_{\bf p+q})\times\nonumber\\
\times\left\{1+\frac{2iv_k\kappa k}{\omega-\xi_k({\bf p+q})+\xi_k({\bf p})
-\Sigma_{k+1}^R(\varepsilon+\omega\xi_{\bf p+q})
+\Sigma^A_{k+1}(\varepsilon\xi_{\bf p})} 
\right\}J_k^{RA}(\varepsilon{\bf p};\varepsilon+\omega{\bf p+q}) 
\label{Jrec} 
\end{eqnarray}
The ``physical'' vertex 
$J^{RA}(\varepsilon{\bf p};\varepsilon+\omega{\bf p+q})$
is determined as $J^{RA}_{k=0}(\varepsilon{\bf p};\varepsilon+
\omega{\bf p+q})$. Recurrence procedure (\ref{Jrec}) takes into account
{\em all} perturbation theory diagrams for the vertex part.
For $\kappa\to 0\quad (\xi\to\infty)$ (\ref{Jrec}) reduces to the series
studied in Ref. \cite{S74} (cf. also Ref. \cite{Sch}), which can be summed
exactly in analytic form. Standard ``ladder'' approximation corresponds in
our scheme to the case of combinatorial factors $v(k)$ in (\ref{Jrec}) being
equal to 1 \cite{ST91}.

\begin{figure}
\epsfxsize=14cm
\epsfysize=20cm
\epsfbox{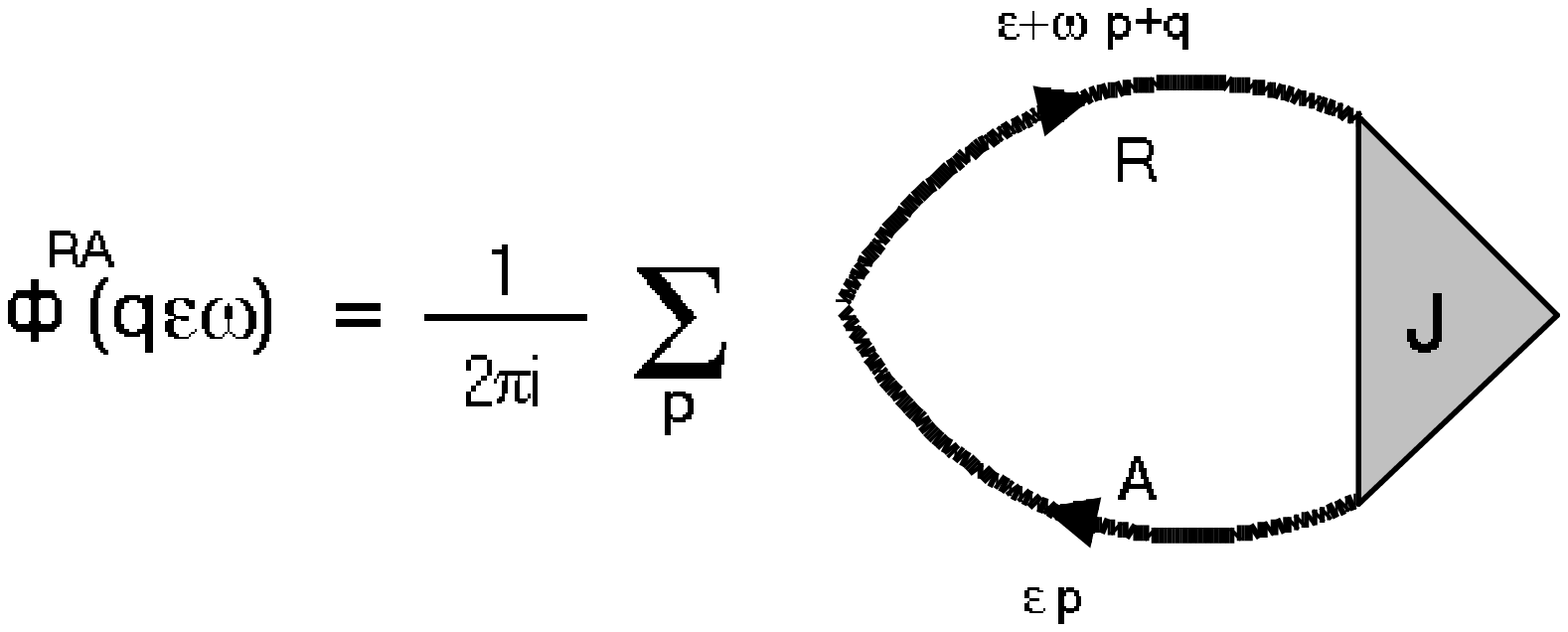}
\caption{Diagrammatic representation for two -- particle response function 
$\Phi^{RA}(q\omega)$.}
\label{loop}
\end{figure}

Conductivity of our system can be expressed \cite{VW} via retarded
density -- density response function $\chi^R(q\omega)$:
\begin{equation}
\sigma(\omega)=e^2\lim_{q\to 0}\left(-\frac{i\omega}{q^2}\right)
\chi^R(q\omega)
\label{conduc}
\end{equation}
where $e$ -- is electronic charge,
\begin{equation}
\chi^R(q\omega)=\omega\left\{\Phi^{RA}(0q\omega)-\Phi^{RA}(00\omega)\right\}
\label{chiR}
\end{equation}
while two -- particle Green's function $\Phi^{RA}(\varepsilon q\omega)$ 
is determined by the loop graph shown in Fig. \ref{loop}.

Direct numerical computations confirm that the recurrence procedure 
(\ref{Jrec}) satisfies an exact relation, which directly follows 
(for $\omega\to 0$) from the Ward identity derived in Ref. \cite{VW}:
\begin{equation}
\Phi^{RA}(00\omega)=-\frac{N(E_F)}{\omega}
\label{Ward}
\end{equation}
where $N(E_F)$ -- is the density of states at the Fermi level $E_F=\mu$. 
This is the main argument for the validity of an {\em Ansatz} used to
derive Eqs. (\ref{J1}),\ (\ref{Jk}) and (\ref{Jrec}).

Finally we can write conductivity in the following symmetrized form,
convenient for numerical computations:
\begin{eqnarray}
\sigma(\omega)=\frac{e^2\omega^2}{\pi}\lim_{q\to 0}\frac{1}
{q^2}\sum_{\bf p}\left\{G^R\left(\frac{\omega}{2},{\bf p}
+\frac{{\bf q}}{2}\right)J^{RA}\left(\frac{\omega}{2},{\bf p}
+\frac{{\bf q}}{2};-\frac{\omega}{2},{\bf p}-\frac{{\bf q}}{2}\right)
G^A\left(-\frac{\omega}{2},{\bf p}-\frac{{\bf q}}{2}\right)-\right.\nonumber\\
\left.-G^R\left(\frac{\omega}{2},{\bf p}\right)
J^{RA}\left(\frac{\omega}{2},{\bf p};-\frac{\omega}{2},{\bf p}\right)
G^A\left(-\frac{\omega}{2},{\bf p}\right)\right\}\nonumber\\ 
\label{optcond} 
\end{eqnarray}
where we have introduced an additional factor of 2 due to spin summation.

Direct numerical computations were performed using (\ref{optcond}),\
(\ref{Jrec}),\ (\ref{G}), with recurrence procedure being cut for high enough
$k$, where all $\Sigma_k$ are $J_k$ were supposed to be equal to zero.
Integration in (\ref{optcond}) was made over the whole Brillouin zone.
The ``bare'' electronic spectrum was taken from (\ref{spectr}). 
Integration momenta are made dimensionless in a natural way with the help of
the lattice constant $a$, while all energies in what follows are in 
units of the transfer integral $t$. Conductivity is measured in units of
the universal conductivity in two -- dimensions 
$\sigma_0=\frac{e^2}{\hbar}=2.5\ 10^{-4}$ Ohm$^{-1}$, and the density of states
--- in units of $1/ta^2$.

\section{Results and Discussion.}

\begin{figure}
\epsfxsize=15cm
\epsfysize=19cm
\epsfbox{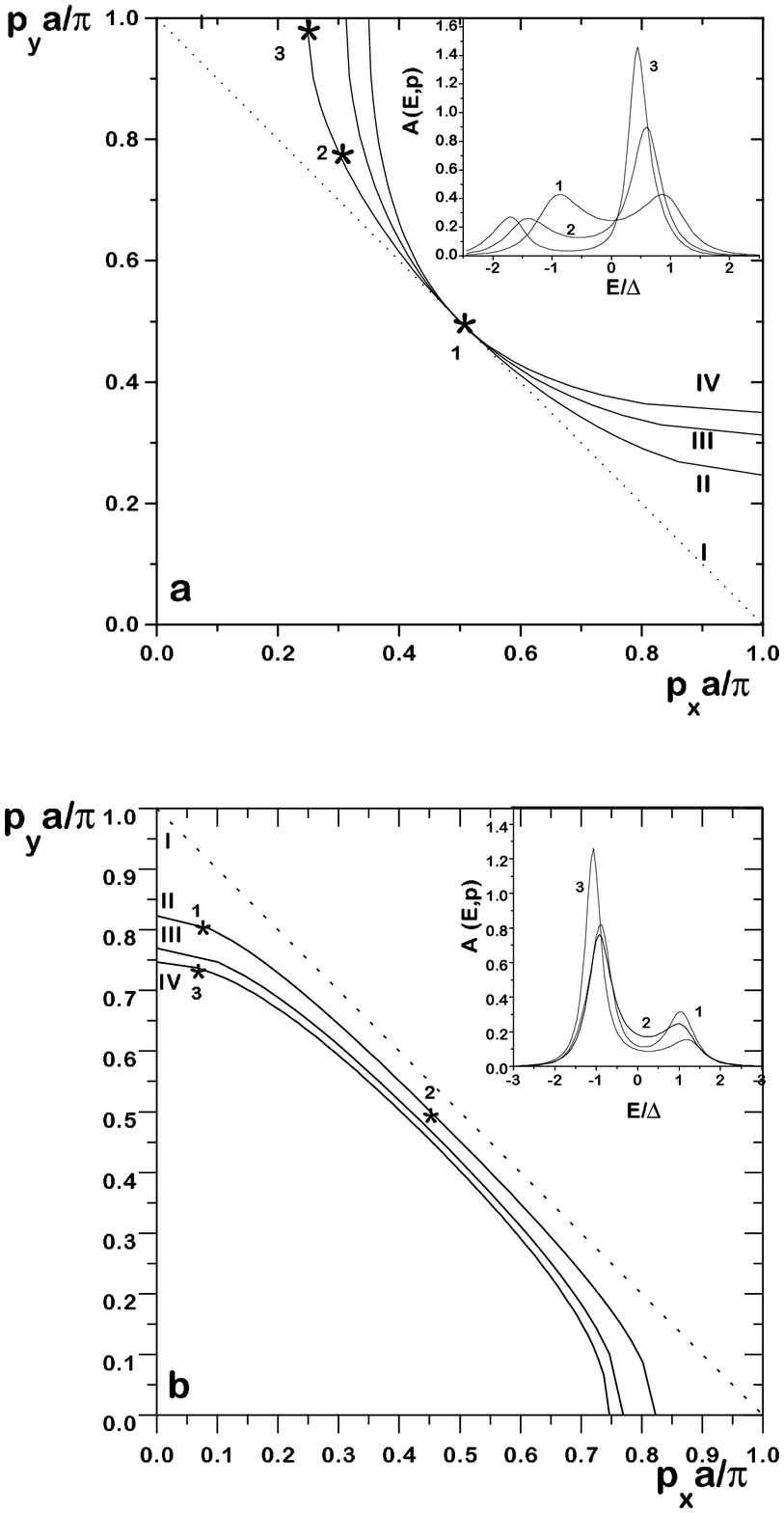}
\caption{Fermi surfaces for different values of $t'$ and chemical potential 
$\mu$.
(a) corresponds to $\mu=0$ and the following values of $t'/t$:\
0 -- I;\ -0.2 -- II;\ -0.4 -- III;\ -0.6 -- IV.
(b) corresponds to  $t'=0$ and following values of $\mu/t$:
0 -- I;\ -0.3 -- II;\ -0.5 -- III;\ -0.6 -- IV.
At the inserts -- energy dependences of the spectral density for spin -- 
fermion model for $\kappa a=0.1$ at the points in the momentum space denoted
by stars.}
\label{FSsqr}
\end{figure}

\begin{figure}
\epsfxsize=14cm
\epsfysize=14cm
\epsfbox{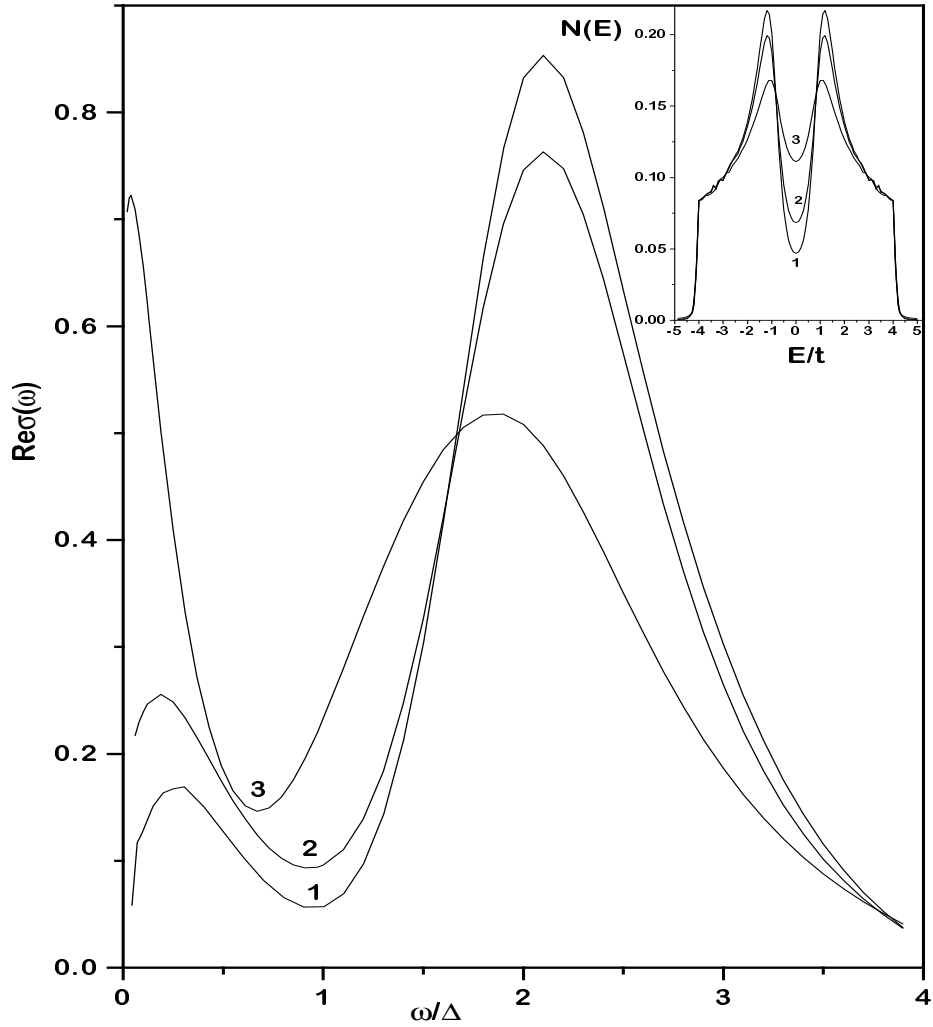}
\caption{Real parts of optical conductivity in spin -- fermion model
for the case of square Fermi surface 
($\mu=0$, $t'=0$) for different values of inverse correlation length
of short -- range order ${\kappa a}$:\
0.1 -- 1;\ 0.2 -- 2;\ 0.5 -- 3.
At the insert -- appropriate densities of states.}
\label{condsqr}
\end{figure}

Optical conductivity and some other characteristics of the model were
computed for different values of parameters, determining the ``bare''
quasiparticle spectrum (\ref{spectr}) and for fixed value of $\Delta=t$.
Let us consider first Fermi surfaces close to the case of half -- filled band 
with $\mu=0$ and $t'=0$, shown (for the first quadrant of the Brillouin zone) 
in Fig. \ref{FSsqr} (a). For  $\mu=0$ and $t'=0$ the Fermi surface is a
well known simple square (complete nesting), so that this case is quite
similar to that of one -- dimensional system, considered in Refs.
\cite{S74,S91,ST91}. Our results for the real part of optical conductivity
of two -- dimensional model, for the case of spin -- fermion combinatorics of
diagrams and different values of correlation length of AFM short -- range
order (parameter $\kappa=\xi^{-1}$, where $\xi$ is measured in units of lattice
spacing $a$) are shown in Fig. \ref{condsqr}. Qualitative form of conductivity
is similar to that found previously for the one -- dimensional model
(for the case of incommensurate CDW -- fluctuations) in Refs. \cite{S91,ST91}.
It is characterized by a sharp maximum due to pseudogap absorption
(corresponding densities of states, demonstrating pseudogap at the Fermi
level, are shown at the insert in Fig. \ref{condsqr}) at
$\omega\sim 2\Delta$ and also by weaker maximum at smaller frequencies due to
Anderson localization of carriers in the random field of static AFM
fluctuations. Localization nature of this maximum is confirmed by its
transformation into the usual ``Drude -- like'' peak 
(with maximum at $\omega=0$) if we perform calculations in the 
``ladder'' approximation, when combinatorial factors $v(k)=1$, corresponding to
``switching off'' the contribution of diagrams with intersecting interaction
lines, leading to two -- dimensional Anderson localization \cite{VW,GLK}.
Qualitative form of conductivity in this case is also quite similar to that
obtained previously for one -- dimensional model in Ref. \cite{ST91}. 
The narrowing of localization peak with diminishing correlation length of
fluctuations can be explained, as was noted in Ref. \cite{ST91}, by
weakening of effective interaction (\ref{Veff}) for smaller values of $\xi$
(for the fixed value of $\Delta$), leading to general weakening of 
scattering rate also at the ``cold'' parts of the Fermi surface.
Note that general behavior of our density of states and optical conductivity
is in full qualitative agreement with the results of quantum
Monte -- Carlo calculations of a similar model of two -- dimensional
Peierls transition obtained recently in Ref. \cite{ScM}.

\begin{figure}
\epsfxsize=14cm
\epsfysize=14cm
\epsfbox{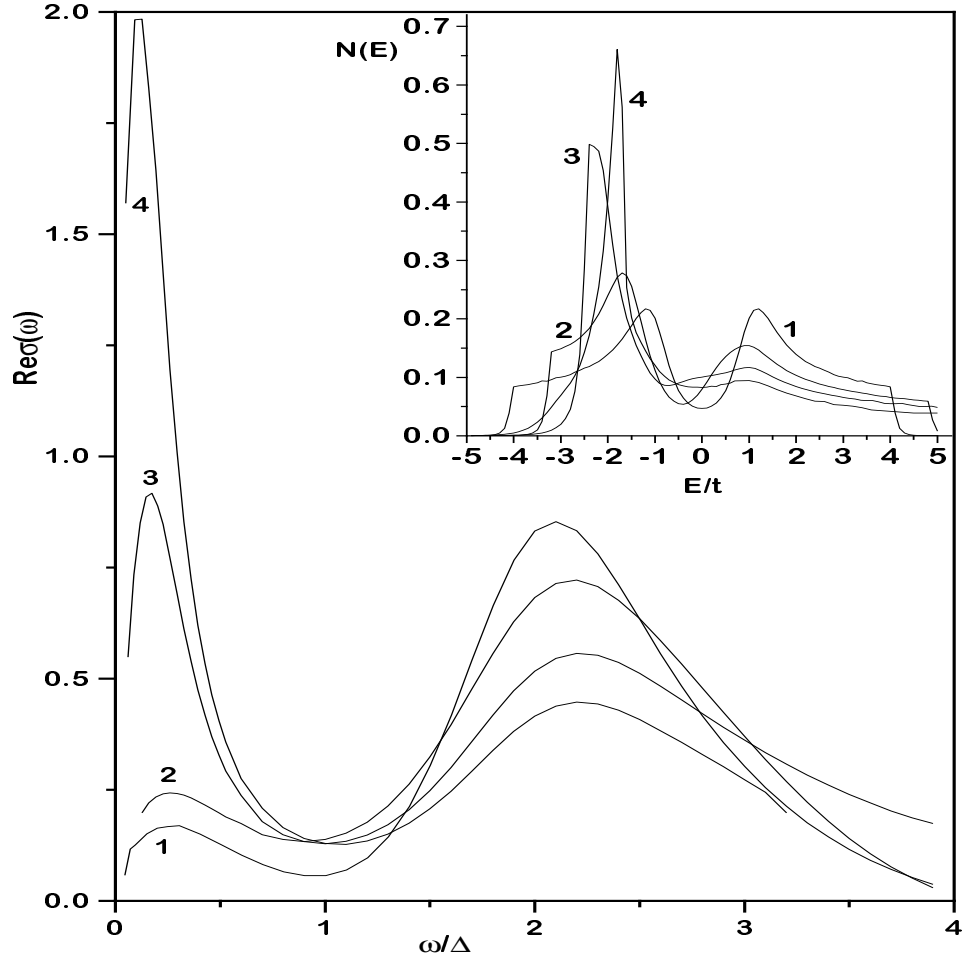}
\caption{Real part of optical conductivity in spin -- fermion model for 
$\mu=0$ and ${\kappa a}=0.1$  for different Fermi surfaces, obtained from the
square by ``switching on'' the transfer integral $t'/t$:\
0 -- 1;\ -0.2 -- 2;\ -0.4 -- 3;\ -0.6 -- 4.
At the insert -- appropriate densities of states.}
\label{condFSt}
\end{figure}

Consider now the case of $\mu=0$ again, but ``switch on'' the second nearest
neighbor transfer integral $t'$ in (\ref{spectr}). Then we are dealing
with Fermi surfaces different from the square one, as shown in 
Fig. \ref{FSsqr} (a). At the insert in this figure we show the energy
dependence of spectral densities (\ref{spdns}) in several typical points of
these Fermi surfaces. It can be seen that these spectral densities have
characteristic  ``non -- Fermi liquid'' like behavior of the type studied
in Refs. \cite{Sch,KS}, practically everywhere at the Fermi surface, while
this surface is not very far from the square, despite the fact that the
``hot spot'' in this case is precisely at the crossing of the zone diagonal
with the Fermi surface. Corresponding frequency dependences of the real part
of optical conductivity are shown in Fig. \ref{condFSt}. At the insert in
this figure we show corresponding densities of states. It is seen that as
we depart from the situation of complete nesting, conductivity maximum due
to pseudogap transitions becomes smaller, while the localization peak grows
and becomes sharper (in accordance with the general sum rule for conductivity). 
Note, however, that pseudogap absorption peak remains noticeable even in the
case, when the pseudogap in the density of states is practically absent
(curves 4 in Fig. \ref{condFSt}). 

\begin{figure}
\epsfxsize=14cm
\epsfysize=14cm
\epsfbox{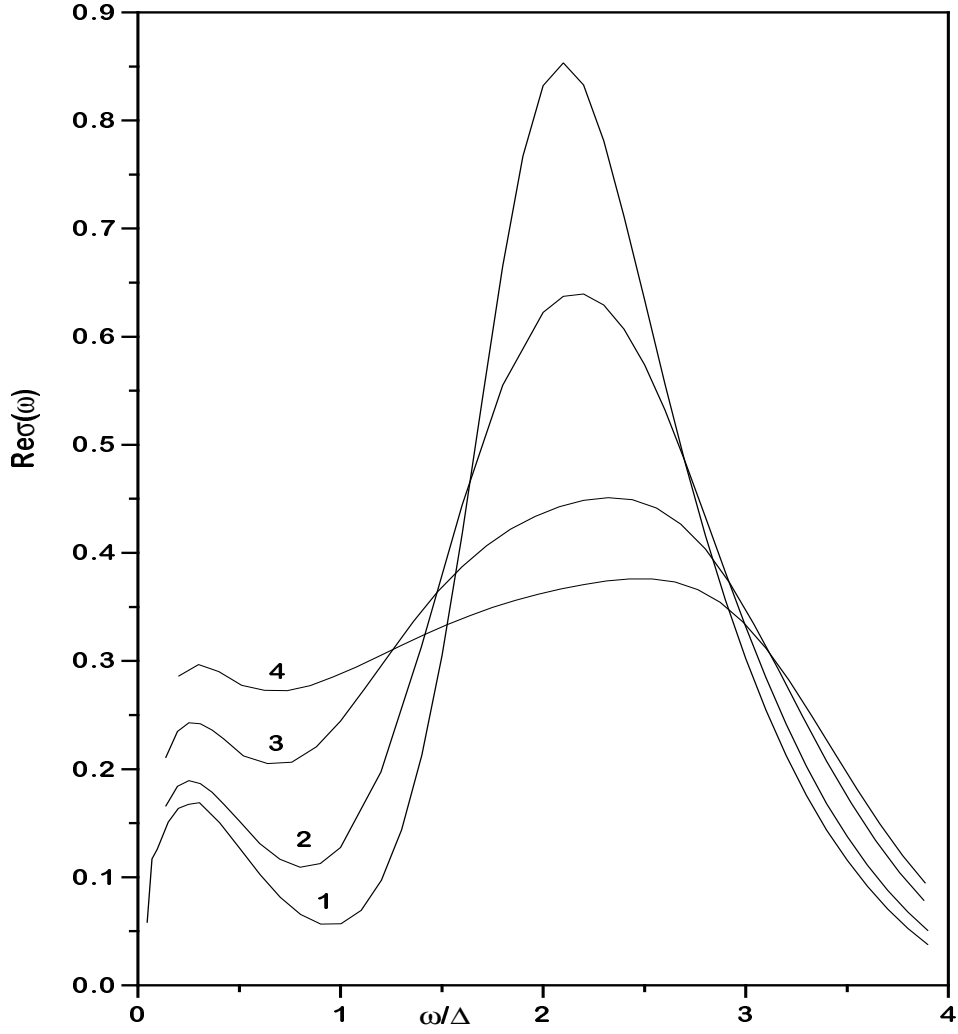}
\caption{Real part of optical conductivity in spin -- fermion model for
$t'=0$ and ${\kappa a}=0.1$ for different Fermi surfaces, obtained from the
square by the shift from  half -- filled band. Chemical potentials correspond
to the following values of $\mu/t$:\
0 -- 1;\ -0.3 -- 2;\ -0.5 -- 3;\ -0.6 -- 4.}
\label{condFSm}
\end{figure}

Let us now return to the case of $t'=0$ and vary chemical potential 
$\mu$, obtaining Fermi surfaces, which are sufficiently close to square,
as shown in Fig. \ref{FSsqr} (b). Strictly speaking there are no 
``hot spots'' at these Fermi surfaces at all, but the spectral density,
shown at the insert in Fig. \ref{FSsqr} (b) still conserves a typical pseudogap
form. Corresponding dependences for the real part of optical conductivity are
shown in Fig. \ref{condFSm}. 

\begin{figure}
\epsfxsize=15cm
\epsfysize=19cm
\epsfbox{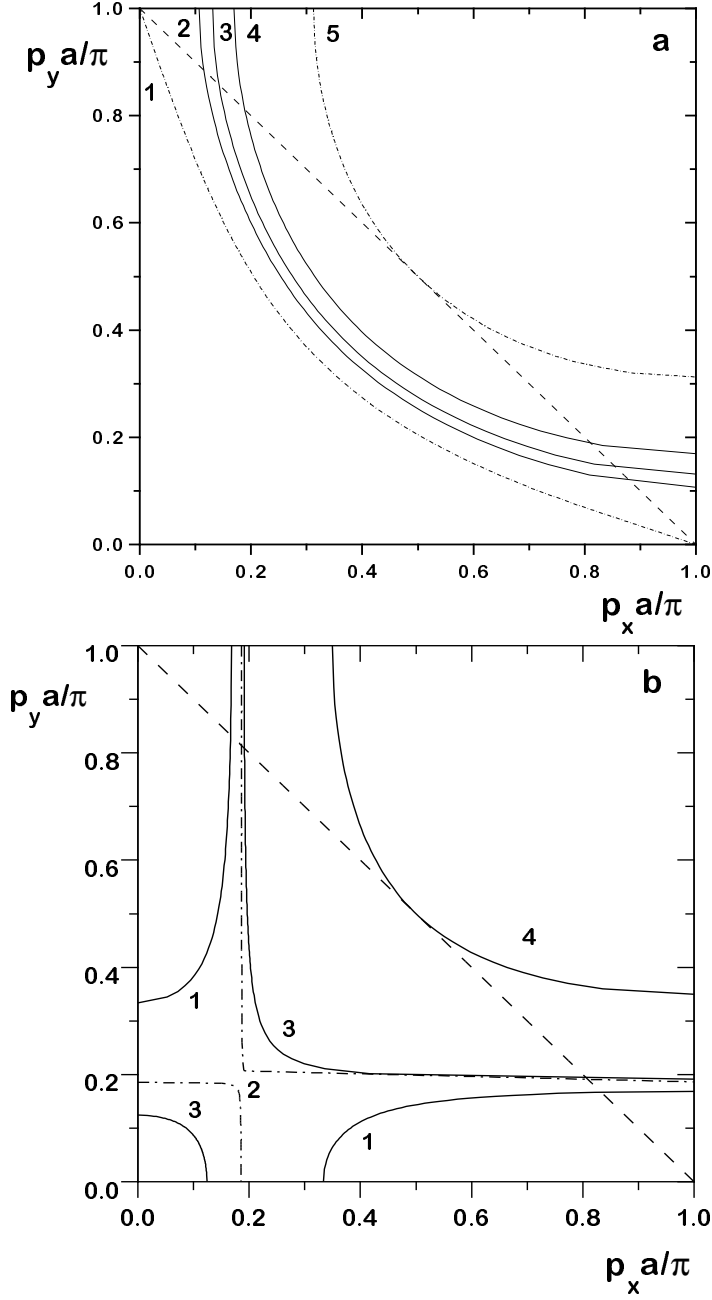}
\caption{
Fermi surfaces for different values of $t'$ and chemical potential $\mu$.
(a) corresponds to the case of $t'/t=-0.4$, typical for HTSC -- cuprates,
and the following values of $\mu/t$:\
-1.6 -- 1;\ -1.4 -- 2;\ -1.3 -- 3;\ -1.1 -- 4;\ 0 -- 5.
``Hot spots'' exist for $-1.6<\mu/t<0$.
(b) corresponds to the case of $t'/t=-0.6$ and the following values of $\mu/t$:\
-1.8 -- 1;\ -1.666 -- 2;\ -1.63 -- 3;\ 0 -- 4.
``Hot spots'' exist for $\mu<0$.
Dashed line -- magnetic Brillouin zone.} 
\label{FSgen} 
\end{figure}

\begin{figure}
\epsfxsize=14cm
\epsfysize=14cm
\epsfbox{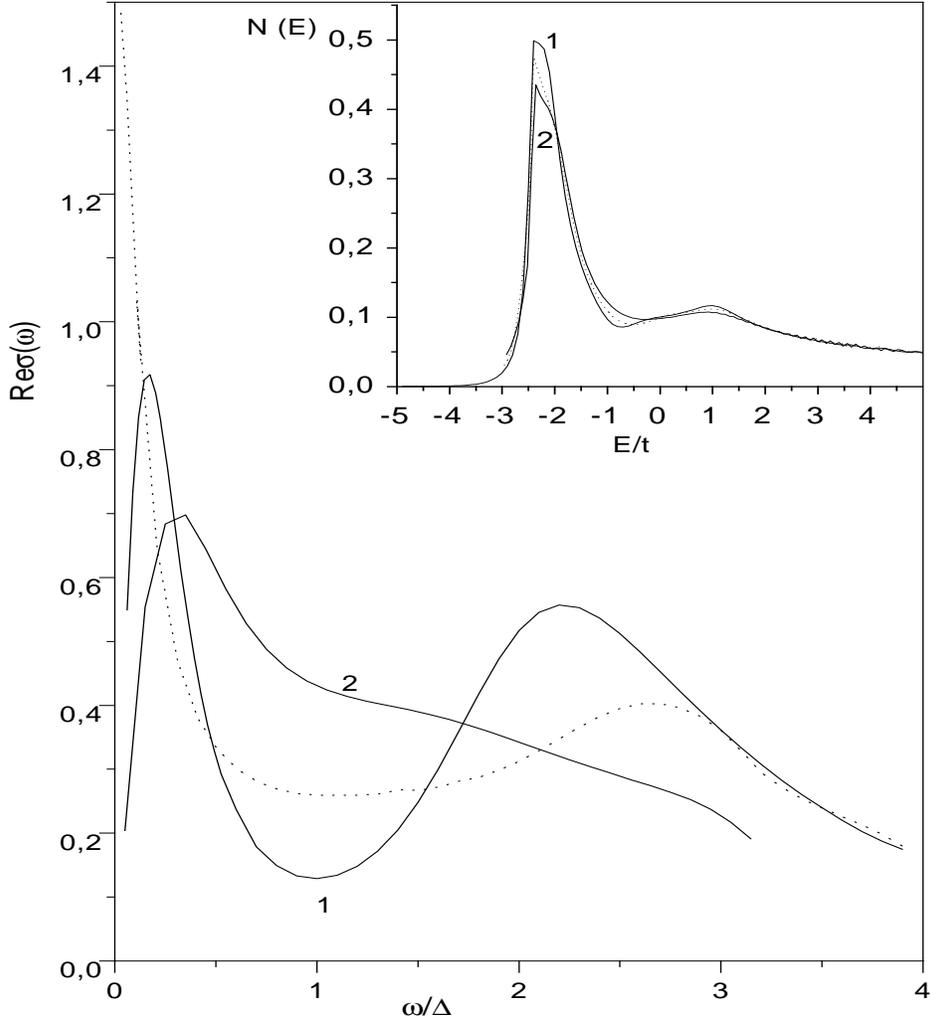}
\caption{Real part of optical conductivity for
$t'/t=-0.4$ and $\mu/t=0$
for ${\kappa a}=0.1$ and different combinatorics of diagrams:\
1 -- spin -- fermion combinatorics;\
2 -- commensurate case.\ Dashed line - ``ladder'' approximation. 
At the insert -- corresponding densities of states.}
\label{condFSga}
\end{figure}

\begin{figure}
\epsfxsize=14cm
\epsfysize=14cm
\epsfbox{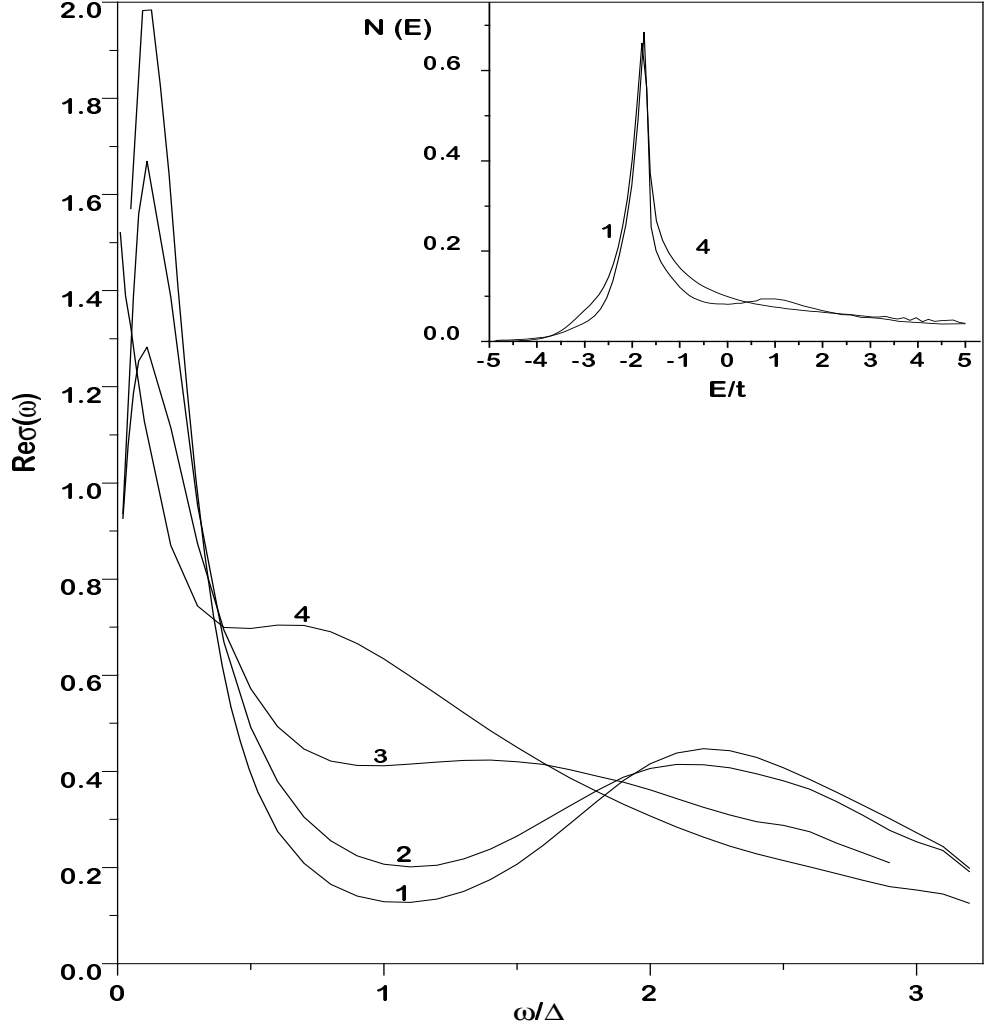}
\caption{Real part of optical conductivity in spin -- fermion model for the
case of $t'/t=-0.6$ and $\mu=0$ for different values of inverse correlation
length ${\kappa a}$:\
0.1 -- 1;\ 0.2 -- 2;\ 0.5 -- 3;\ 1 -- 4.
At the insert -- densities of state, corresponding to 1 and 4.}
\label{condFSgb}
\end{figure}

Consider now different geometries of the Fermi surface with ``hot spots'',
shown in Fig. \ref{FSgen}. In Figs. \ref{condFSga},\ \ref{condFSgb} we show
the real part of optical conductivity, calculated (for different diagram
combinatorics) for two characteristic values of $t'=-0.4t$ and $t'=-0.6t$ 
for the chemical potential value $\mu=0$, when ``hot spots'' are on the zone
diagonal (curve 5 in Fig. \ref{FSgen} (a) and curve 4 in Fig. \ref{FSgen} (b)). 
We can see  again that the pseudogap behavior of conductivity persists even 
for the case, when the pseudogap in the density of states (shown at the inserts
in Figs. \ref{condFSga},\  \ref{condFSgb}) is practically absent. Dashed curve
in Fig. \ref{condFSga} shows the results of the ``ladder'' approximation,
demonstrating typical vanishing of two -- dimensional localization in this
approximation. The results shown in Fig. \ref{condFSgb} demonstrate typical
``smearing'' of the pseudogap maximum of conductivity with diminishing
correlation length of short -- range order fluctuations.

\begin{figure}
\epsfxsize=14cm
\epsfysize=14cm
\epsfbox{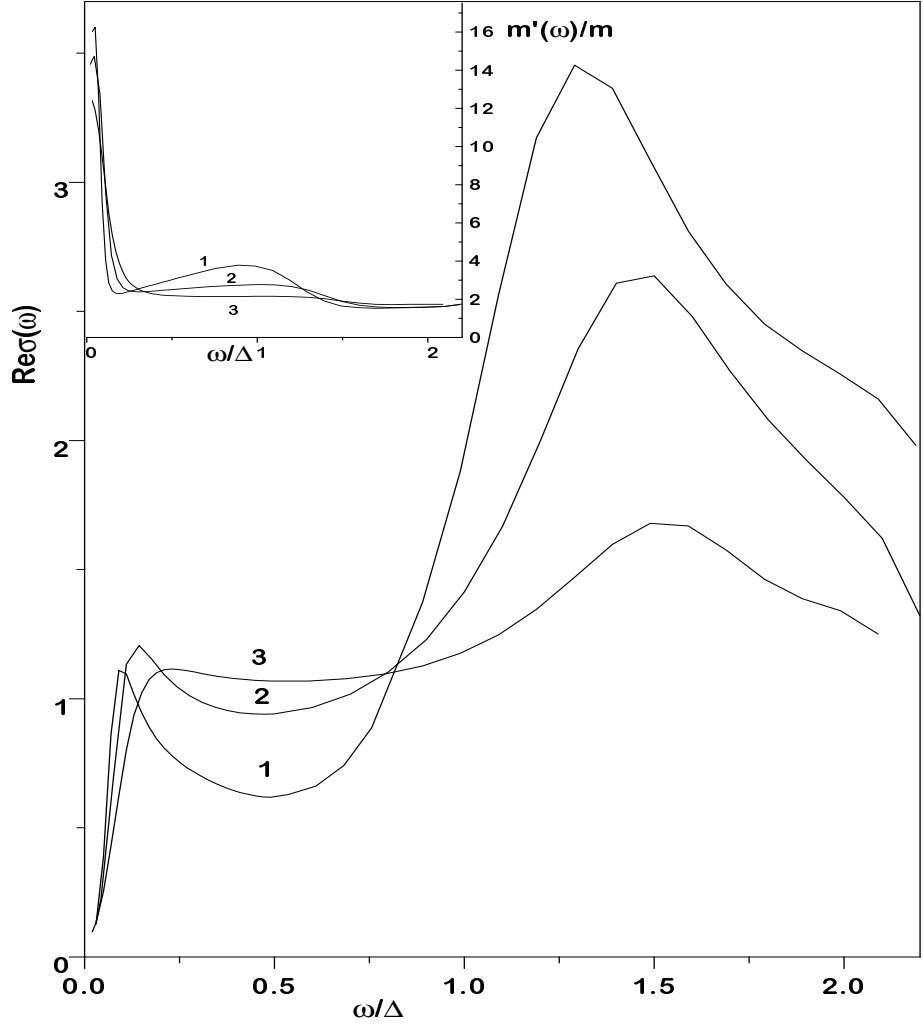}
\caption{Generalized scattering rate and effective mass for the case of
$t'/t=-0.4$ and $\mu/t=-1.3$, typical for high -- temperature superconductors.
Parameters of the generalized Drude model were obtained for spin -- fermion
model with different values of correlation length
${\kappa a}$:\ 0.05 -- 1;\ 0.1 -- 2;\ 0.2 -- 3.\  
Dephasing rate ${\gamma/t}=0.005$.}
\label{tau}
\end{figure}

\begin{figure}
\epsfxsize=14cm
\epsfysize=14cm
\epsfbox{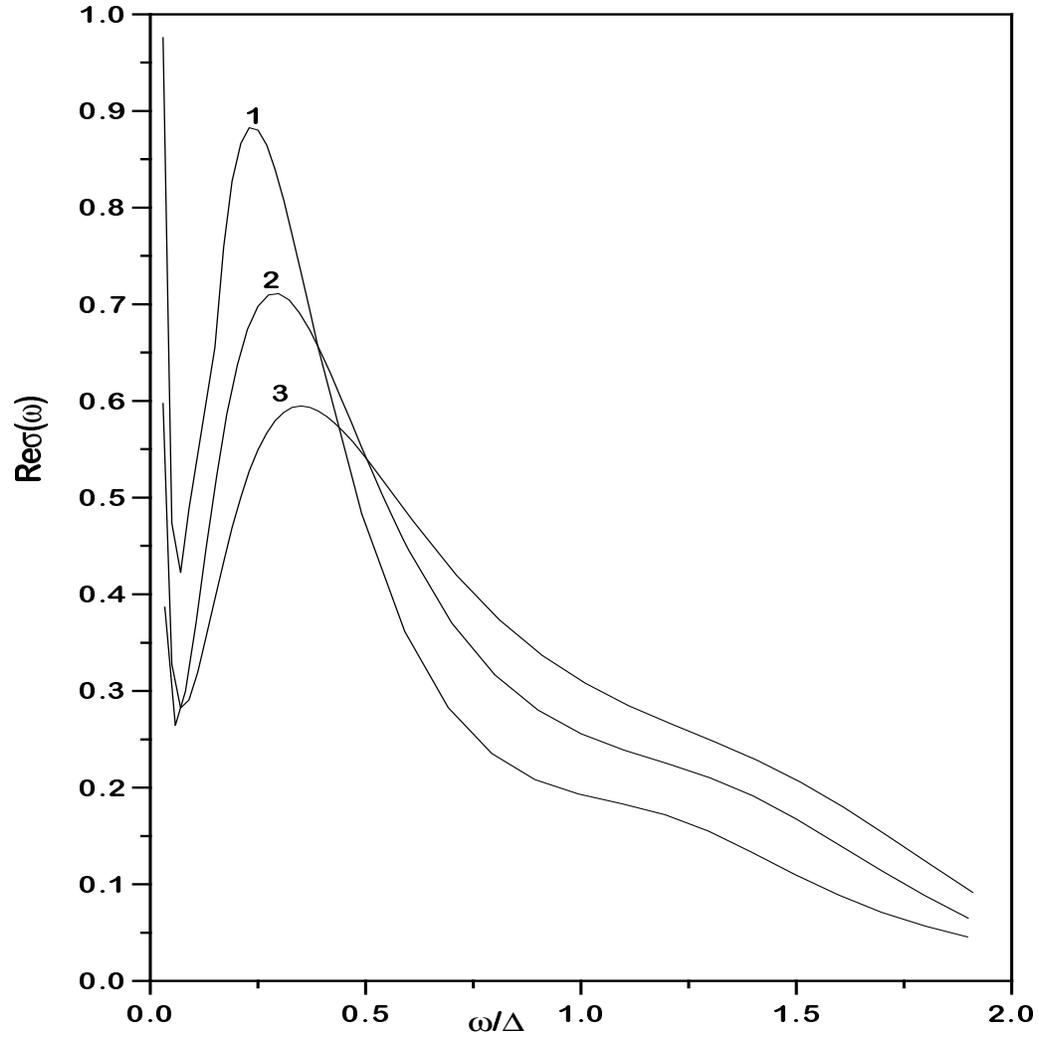}
\caption{Real part of optical conductivity for commensurate case with 
$t'/t=-0.4$ and $\mu/t=-1.3$ and different values of inverse correlation
length ${\kappa a}$:\ 0.05 -- 1;\ 0.1 -- 2;\ 0.2 -- 3.  
Dephasing rate ${\gamma}=0.005/t$.}
\label{condHSc}
\end{figure}

For majority of copper oxide high -- temperature superconductors typical 
geometry of the Fermi surface is  described by the case of $t'=-0.4t$ and
$\mu=-1.3t$ \cite{Sch} (as shown by curve 3 in Fig. \ref{FSgen} (a)). Results
of our calculations of optical conductivity for this case and for different
values of the inverse correlation length $\kappa$ are shown in 
Fig. \ref{condHS} (for spin -- fermion combinatorics of diagrams). Here we
also introduced an additional weak scattering due to inelastic processes,
using the standard substitution $\omega\to\omega+i\gamma$ \cite{GZ}, 
which leads to the appearance of a narrow ``Drude -- like'' peak in the
region of $\omega < \gamma$ (destruction of two -- dimensional localization
due to dephasing). It can be easily checked that for higher values of
inelastic scattering rate $\gamma$ localization peak is ``smeared'' and
transforms into the ``usual'' Drude -- like peak in the region of small
frequencies. Pseudogap absorption maximum becomes more pronounced with the
growth of correlation length $\xi$ (diminishing $\kappa$). 
In Fig. \ref{tau} we show frequency dependencies of an effective scattering
rate $1/\tau(\omega)$ and effective mass $m^*(\omega)$, which can be 
determined from our calculations, using the generalized Drude formula,
which is often used to fit experimental data \cite{Tim}:
\begin{eqnarray}
\frac{1}{\tau(\omega)}=\frac{\omega_p^2}{4\pi}Re\left(\frac{1}{\sigma(\omega)}
\right) \label{sgdr}\\
\frac{m^*(\omega)}{m}=-\frac{1}{\omega}\frac{\omega^2_p}{4\pi}Im\left(
\frac{1}{\sigma(\omega)}\right)
\label{mgdr}
\end{eqnarray}
Here $\omega_p$ -- is plasma frequency, $m$ -- is the mass of a free electron.
From Fig. \ref{tau} we can see, that $1/\tau(\omega)$ (expressed in this figure
in units of $\frac{\omega^2_p}{4\pi e^2}\hbar$) demonstrates quite typical
pseudogap behavior for the frequency range of $\omega < 2\Delta$. Note, that
the density of states in this case possess only rather weak pseudogap
\cite{KS} (cf. insert in Fig. \ref{condFSga}). In Fig. \ref{condHSc} we show
similar results for the same case (typical for HTSC -- oxides), but 
for commensurate combinatorics of diagrams (CDW -- type fluctuations).
We see that in this case the pseudogap absorption maximum is almost
invisible.

\begin{figure}
\epsfxsize=14cm
\epsfysize=14cm
\epsfbox{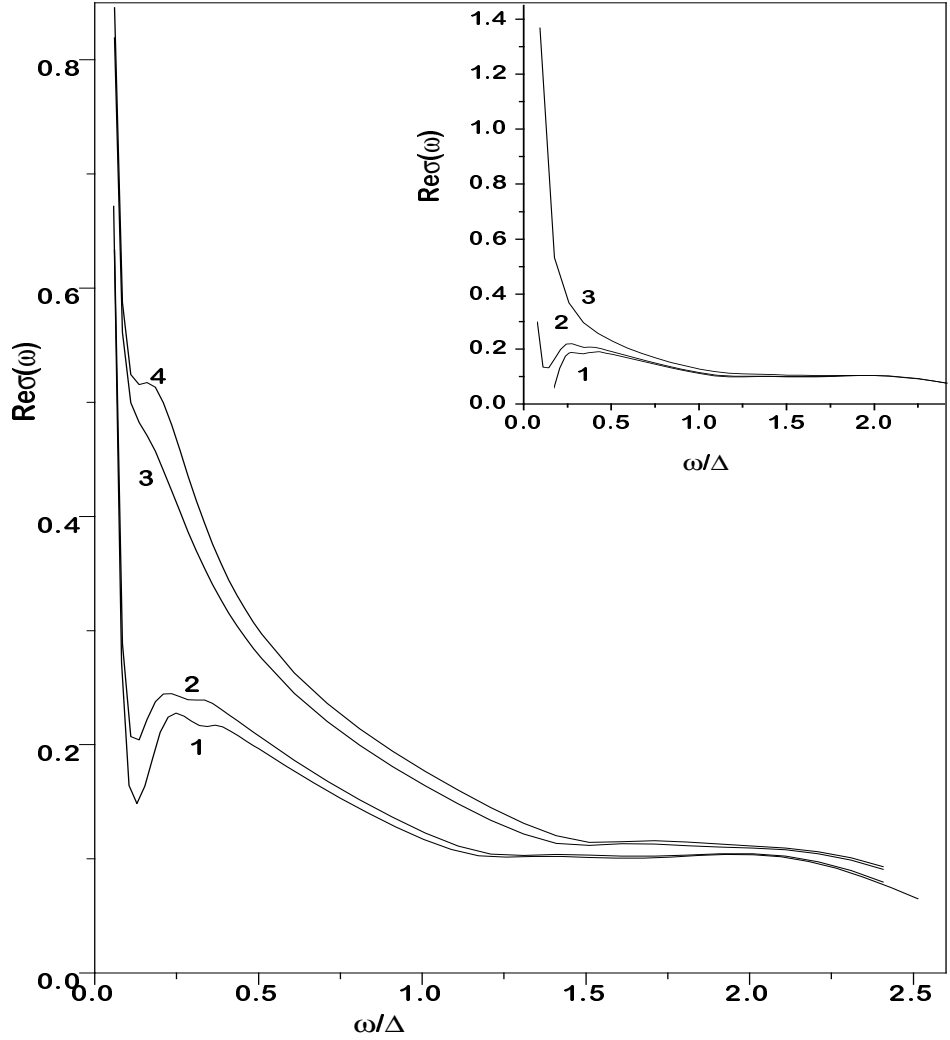}
\caption{Evolution of the real part of optical conductivity 
for commensurate case with $t'/t=-0.6$ and ${\kappa a}=0.2$ and chemical
potential changing in the vicinity of topological transition.
Different curves correspond to the following values of $\mu/t$:\
-1.79 -- 1;\ -1,77 -- 2;\ -1.66 -- 3;\ -1.63 -- 4.
Dephasing rate ${\gamma/t}=0.01$.
At the insert -- real part of optical conductivity for the case of
$\mu/t=-1.8$ for different values of ${\gamma/t}$:\ 
0 -- 1;\ 0.01 -- 2;\ 0.05 -- 3.}
\label{condtop}
\end{figure}

From Fig. \ref{FSgen} (b) we can see that with the change of the chemical
potential in the interval from $\mu=0$ to $\mu=-1.666t$ the Fermi surface
acquires larger and larger ``flat'' parts, transforming for
$\mu\approx 1.666t$ practically into a ``cross -- like''. Similar Fermi
surface was observed in ARPES -- experiments on $La_{1.28}Nd_{0.6}
Sr_{0.12}CuO_4$ \cite{Zh,DLS}. In this case velocity projections in
``hot spots'', connected by ${\bf Q}=(\frac{\pi}{a},\frac{\pi}{a})$, become
orthogonal. For $\mu/t=-1.666...$ topology of the Fermi surface change
(cf. Fig. \ref{FSgen} (b)) and in the whole region of $\mu/t<-1.666...$ these
projections have the same signs, guaranteeing the exactness of our main
{\em Ansatz} (\ref{Ansatz}) for the contributions of higher order diagrams
\cite{KS}. It is of some interest to present the results of our calculations
of optical conductivity also for these values of $\mu$. These are shown (for
the case of commensurate (CDW) diagram combinatorics) in Fig. \ref{condtop},
where we can follow the changes of localization peak in conductivity as
the chemical potential passes through the region of topological phase
transition. A weak pseudogap absorption maximum practically does not change
at all. At the insert in Fig. \ref{condtop} we show the evolution of
localization peak as we ``switch on'' inelastic scattering processes
(scattering rate $\gamma$) for the case of $\mu=-1.8t$. The transition from
localization to Drude -- like behavior due to inelastic dephasing is clearly
seen. These results show, that the change in topology of the Fermi surface
does not lead to any strong qualitative change of optical conductivity in
our model.

\section{Conclusion}

Our analysis shows the variety of results, which may be obtained in this
model for different geometries and topologies of the Fermi surface, appearing 
for different values of the ``bare'' quasiparticle energy spectrum
(\ref{spectr}). It is interesting to compare these data with results obtained
earlier for the simplified model with ``hot patches'' on the Fermi surface
\cite{S99}. Pseudogap anomalies in the ``hot patches'' model were determined
mainly by strong scattering on these (flat) patches only, as well as by their
relative size. Accordingly, the localization peak of conductivity in this
model was rather weak and dominating behavior at small frequencies was
determined by Drude -- like peak due to the scattering on ``cold'' parts of the
Fermi surface with the scattering rate $\gamma$ (analogous to inelastic
scattering rate defined above). Analysis of the more realistic model, given
in this paper, shows that the contribution of localization peak may be more
pronounced and in fact this peak may transform into a narrow enough 
``Drude -- like'' peak with the inclusion of dephasing processes.

Probably the main deficiency of the present model is the neglect of the
dynamics of fluctuations of short -- range order. This approximation is
justified, as was noted in Refs. \cite{Sch,KS}, only for high enough
temperatures. However, at higher temperatures inelastic scattering, responsible
for dephasing processes and breakdown of localization, become more important.
The other shortcoming, as was stressed several times \cite{S79,KS}, 
is our limitation to Gaussian fluctuations only of short -- range order, 
which also may be justified only for high enough temperatures.

\begin{figure}
\epsfxsize=14cm
\epsfysize=14cm
\epsfbox{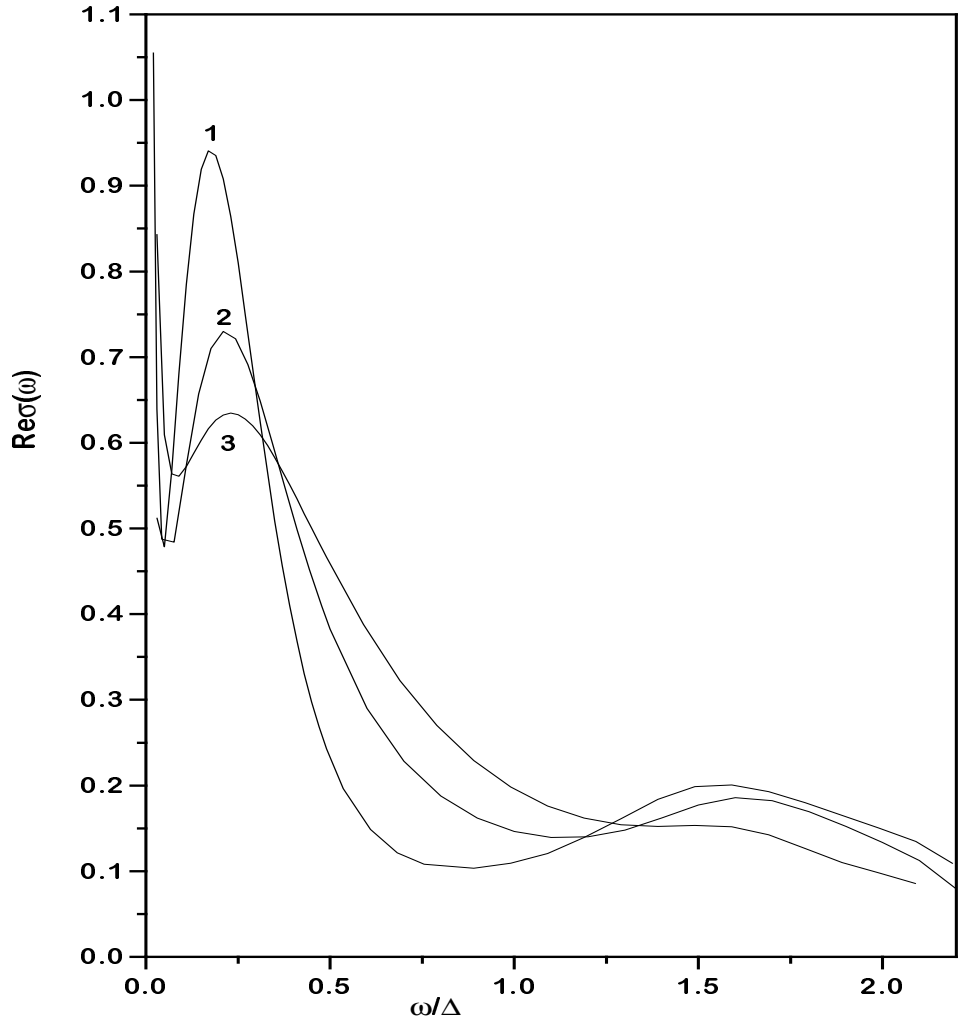}
\caption{Real part of optical conductivity in spin -- fermion model for 
$t'/t=-0.4$ and $\mu/t=-1.3$ and different values of correlation length
${\kappa a}$:\ 0.05 -- 1;\ 0.1 -- 2;\ 0.2 -- 3.\  
Dephasing rate ${\gamma/t}=0.005$.}
\label{condHS}
\end{figure}

Discussing the possible relation of the above results with experimental data 
obtained on real HTSC -- cuprates we note, that in majority of these
experiments \cite{Tim,MS} localization peak was not observed at all, which may
be due to the large contribution of inelastic processes (dephasing) at high
enough temperatures used in these measurements. Peaks in optical conductivity
at small frequencies, attributed to localization, were observed in disordered
samples of $YBaCuO$ in Refs. \cite{B1,B2}. In recent experiments on $NdCeCuO$ 
\cite{B3,Ono} this peak was especially clearly seen. In particular, the 
qualitative behavior of optical conductivity observed in Ref. \cite{Ono}
for several samples of $NdCeCuO$ with different compositions
(from underdoped to optimally doped) is in complete agreement with our
results shown above in Fig. \ref{condHS}, which were considered as typical
for HTSC -- cuprates. We conclude that the ``hot spots'' model may 
be successfully used for realistic enough description of anomalies of
optical conductivity in high -- temperature superconductors.

The authors are grateful to E.Z.Kuchinskii for numerous discussions.
This work was supported in part by the grants of the Russian Foundation of
Basic Research 02-02-16031, CRDF No. REC-005 and the program of fundamental 
research of the Presidium of the Russian Academy of Sciences ``Quantum
Macrophysics''. It was also partly supported by the project of the Russian
Ministry of Industry and Science ``Studies of collective and quantum
effects in condensed matter''.

\newpage

\newpage

\newpage

\newpage

\newpage

\newpage

\newpage

\newpage

\newpage

\newpage

\newpage

\newpage

\newpage

\newpage

\newpage

\newpage

\newpage

\end{document}